\documentclass[aps,preprint,showpacs,floatfix]{revtex4}
\usepackage{graphicx,bm,amsmath,dcolumn,}
\usepackage{verbatim}
\usepackage{color}
\usepackage{epstopdf}
\usepackage{chngcntr}

\begin{document}
\newcommand{\be}{\begin{equation}}
\newcommand{\ee}{\end{equation}}
\newcommand{\half}{\frac{1}{2}}
\newcommand{\ith}{^{(i)}}
\newcommand{\im}{^{(i-1)}}
\newcommand{\gae}
{\,\hbox{\lower0.5ex\hbox{$\sim$}\llap{\raise0.5ex\hbox{$>$}}}\,}
\newcommand{\lae}
{\,\hbox{\lower0.5ex\hbox{$\sim$}\llap{\raise0.5ex\hbox{$<$}}}\,}

\definecolor{blue}{rgb}{0,0,1}
\definecolor{red}{rgb}{1,0,0}
\definecolor{green}{rgb}{0,1,0}
\newcommand{\blue}[1]{\textcolor{blue}{#1}}
\newcommand{\red}[1]{\textcolor{red}{#1}}
\newcommand{\green}[1]{\textcolor{green}{#1}}

\newcommand{\gn}{g_{\rm n}}
\newcommand{\gnz}{g_{\rm n,0}}
\newcommand{\gnn}{g_{\rm nn}}
\newcommand{\gnnz}{g_{\rm nn,0}}
\newcommand{\Cn}{C_{\rm n}}
\newcommand{\Cnz}{C_{\rm n, 0}}
\newcommand{\Cnn}{C_{\rm nn}}
\newcommand{\Cnnz}{C_{\rm nn,0}}
\newcommand{\pctri}{p_c^{\rm tri}}

\newcommand{\calA}{ {\mathcal A}}
\newcommand{\calC}{ {\mathcal C}}
\newcommand{\calE}{ {\mathcal E}}
\newcommand{\calG}{ {\mathcal G}}
\newcommand{\calH}{ {\mathcal H}}
\newcommand{\calO}{ {\mathcal O}}
\newcommand{\calS}{ {\mathcal S}}
\newcommand{\calZ}{ {\mathcal Z}}
\newcommand{\calN}{ {\mathcal N}}
\title{Short-range correlations in percolation at criticality}
\author{Hao Hu$^1$, Henk W. J. Bl\"ote$^2$, 
Robert M. Ziff$^3$, and Youjin Deng$^1$\footnote{yjdeng@ustc.edu.cn}}
\affiliation{$^{1}$ Hefei National Laboratory for Physical
Sciences at Microscale,
Department of Modern Physics, University of Science and
Technology of China, Hefei 230027, China }
\affiliation{$^{2}$ Instituut Lorentz, Leiden University,
P.O. Box 9506, 2300 RA Leiden, The Netherlands}
\affiliation{$^{3}$ Center for the Study of Complex Systems 
and Department of Chemical Engineering,
University of Michigan, Ann Arbor, Michigan 48109-2136 USA}
\date{\today}
\begin{abstract}
We derive the critical nearest-neighbor connectivity $\gn$ as $3/4$, 
${3(7-9\pctri)}/{4(5-4\pctri)}$ and ${3(2+7\pctri})/{4(5-\pctri)}$
for bond percolation on the square, honeycomb and triangular lattice respectively, 
where $\pctri=2\sin(\pi/18)$ is the percolation threshold for the triangular lattice,
and confirm these values via Monte Carlo simulations. 
On the square lattice, we also numerically  determine the critical next-nearest-neighbor
connectivity as $\gnn=0.687\; 500\; 0(2)$, which confirms a conjecture by Mitra and
Nienhuis in J. Stat. Mech. P10006 (2004), implying the exact value $\gnn=11/16$. 
We also determine the connectivity on a free surface as $\gn^{\rm surf}=0.625\; 000\; 1(13)$ 
and conjecture that this value is exactly equal to $5/8$.
In addition, we find that at criticality, the connectivities depend on the linear finite
size $L$ as  $\sim L^{y_t-d}$, and the associated specific-heat-like quantities $\Cn$ and
$\Cnn$ scale as $ \sim L^{2y_t-d} \ln (L/L_0)$, where $d$ is the lattice dimensionality,
$y_t=1/\nu$ the thermal renormalization exponent, and $L_0$ a non-universal constant.
We provide an explanation of this logarithmic factor within the theoretical framework
reported recently by Vasseur et al.~in J. Stat. Mech. L07001 (2012).
\end{abstract}
\pacs{64.60.ah, 68.35.Rh, 11.25.Hf}
\maketitle

\section{Introduction}
\label{intro}
To study the nature of the percolation process \cite{SA}, much attention has been 
paid to correlation functions $P_n (z_1,\cdots,z_n)$ describing the probability 
that $n$ points $(z_1, \cdots, z_n)$ belong to the same cluster. 
For example, the mean cluster size 
can be calculated as $S=\sum_z{P_2(z,0)}$, and a recent work investigated 
the factorization of the three-point correlation function in terms of 
two-point correlations \cite{DVZSK}.  While most results in the literature 
deal with long-range correlations \cite{SA, DVZSK, VJS, VJ14},
the present work is dedicated to the investigation of short-range correlations,  
over distances comparable with the lattice spacing.

It is well known that the bond-percolation model can be considered as 
the $q \rightarrow 1$ limit of the $q$-state Potts model \cite{RBP,FW}.  
For a lattice $G$ with a set of edges denoted as $\{e_{ij}\}$,
the reduced Hamiltonian (i.e., divided by $kT$) of the Potts model reads 
\begin{equation}
  \calH (K,q) = -  K \sum_{e_{ij} } \delta_{\sigma_i \sigma_j} \; ,
  \hspace{10mm} \sigma _i = 1,\cdots, q \; ,
\label{Potts}
\end{equation}
where the sum is over all nearest-neighbor lattice edges $e_{ij}$, and $K=J/kT$,
such that $-J\delta_{\sigma_i \sigma_j}$ is the energy of a neighbor pair.
The celebrated Kasteleyn-Fortuin transformation \cite{KF} maps the 
Potts model onto the random-cluster (RC) model with partition sum
\begin{equation}
Z_{\rm rc}(u,q) = \sum_{\calA \subseteq G} \, u^{\calN_{\rm b}} q^{\calN _{\rm c}} \, , 
\hspace{10mm}  u= e^K-1 \;,
\label{zrc}
\end{equation}
where the sum is over all subgraphs $\calA$ of $G$, 
$\calN_{\rm b}$ is the number of occupied bonds in $\calA$,
and $\calN _{\rm c}$ is the number of connected components (clusters).
The RC model generalizes the Potts model to non-integer values $q >0$, 
and in the limit $q \to 1$ it reduces to the bond-percolation \cite{FW, KF} model, 
in which bonds are uncorrelated, and governed by independent probabilities $p=u/(u+1)$.
As a result, the critical {\em thermal} fluctuations are suppressed in this model,
so that the critical finite-size-scaling (FSS) amplitudes of many
energy-like quantities vanish.  For instance, the density of the occupied bonds 
is independent of the system size, and the density of clusters 
converges rapidly to its background value with zero amplitude for the 
leading finite-size term with $y_t=1/\nu$ in the exponent \cite{ZFA97, HBD12}.
Though the partition function at $q=1$ reduces to a trivial power of $u$,
a number of nontrivial properties of the percolation model can be derived
from the RC model via differentiation of the RC partition sum to $q$, and then
taking the limit $q \to 1$.  Quantities of interest can then be numerically
determined by sampling the resulting expression from Monte Carlo generated 
percolation configurations. An example is given in Appendix~\ref{percsph}, 
where we display the behavior of the RC specific heat in the percolation limit 
$q\to 1$ as a function of temperature.

Making use of existing results on the critical temperature and energy 
of the Potts model \cite{FW,KJ74, BTA78}, in the limit $q\rightarrow1$,  
we derive analytically the critical nearest-neighbor connectivity 
$\gn$ as $3/4$, ${3(7-9\pctri)}/{4(5-4\pctri)}$ and ${3(2+7\pctri})/{4(5-\pctri)}$,
for bond percolation on the square, honeycomb and triangular lattices, respectively, 
where $\pctri=2\sin(\pi/18)$ is the percolation threshold for the triangular lattice,
and confirm them with Monte Carlo simulations.
For bond percolation on the square lattice, we also determine numerically 
the critical next-nearest-neighbor connectivity as $\gnn=0.687\; 500\; 0(2)$ which
is very close to $11/16$. Our transfer-matrix calculations (Appendix~\ref{TMcal}), 
which apply to a cylindrical geometry, are consistent with this value.
As explained in Appendix~\ref{appMapping}, $\gnn$ is related to
a quantity for the completely packed $O(1)$ loop model which has been studied
by Mitra et al.~\cite{MN}. They formulated a conjecture implying the exact
value $\gnn=11/16$. Our results support that this conjecture holds exactly.
Furthermore we determined the connectivity on free one-dimensional surfaces of the
square lattice as $\gn^{\rm surf}=0.625\; 000\; 1(13)$, and
conjecture that this value is exactly equal to $5/8$.

We are also interested in the critical FSS behavior of 
the connectivities $\gn$ and $\gnn$, as well as the associated 
specific-heat-like quantities $\Cn$ and $\Cnn$.
Numerical simulations and finite-size analysis were done for square, 
triangular, honeycomb and simple-cubic lattices. It is found that, at criticality, one has 
$g (L) = g_{\rm a} + g_{\rm s} L^{y_t-d}$ ($d$ is the spatial dimensionality), where 
$g_{\rm a} $ accounts for the background contribution and the amplitude $g_{\rm s}$ for 
the singular part is non-zero. In two and three dimensions, this critical exponent 
is known as $y_t=1/\nu=3/4$ \cite{NDL, CDL} and $y_t=1.141\;0(15)$ \cite{WZZGD}, respectively.
For $\Cn$ and $\Cnn$, it is observed that the leading $y_t$-dependent term with 
exponent 2$y_t-d$ also exists.  Moreover, it is found that this leading term 
is modified by a multiplicative logarithmic factor such that $\Cn$ and $\Cnn$ 
are proportional to $L^{2y_t-d} \ln {(L/L_0)}$, where $L_0$ is non-universal.

The logarithmic factor mentioned above can be related with recently identified
logarithmic observables that were explained by mixing the energy operator 
with an operator connecting two random clusters \cite{VJS, VJ14}. The latter
operator is associated with a change of the bond probability $p$ \cite{Con} 
between Potts spins, while the Potts coupling $K$ remains constant. 
For $q\to 1$ the bond probability field and the temperature field 
become degenerate. This mechanism is independent of the lattice type and 
the number of dimensions.

The remainder of this work is organized as follows. 
Section \ref{OFSS} contains the definitions of the observables, 
as well as their expected FSS behavior.
Section \ref{exact-gn} presents the derivation of the exact critical connectivities.
The Monte Carlo results for $\gn$ and $\gnn$, for $\Cn$ and $\Cnn$,
on different lattices, are presented in Sec.~\ref{Num-results}.
The origin of a logarithmic factor in the FSS behavior of $\Cn$ and $\Cnn$ 
is explored in Sec.~\ref{originLog}.
The paper concludes with a brief discussion in Sec.~\ref{discus}. 
Further details and examples are presented in the Appendices, including the derivation
of the exact nearest-neighbor connectivities for the triangular and honeycomb lattices 
in Appendix \ref{gnHonDer}.

\section{Observables and finite-size scaling}
\label{OFSS}
\subsection{Observables}

We use $\gamma_{x,y}(\calA)=1$ and $0$ to represent the situation that,
in a configuration $\calA$ of bond variables, 
lattice sites $x$ and $y$ belong to the same and to different clusters, respectively.
The following observables were studied:
\begin{enumerate}
\item {\it Energy-like quantities}: 
  \begin{itemize}
    \item The bond-occupation density $\rho_{\rm b} = \langle \calN_{\rm b} \rangle / N_{\rm e}$, 
        where $N_{\rm e}$ denotes the number of edges in the lattice, and ``$\langle \rangle$" 
        represents the ensemble average. 
    \item The cluster-number density $\rho_k =  \langle \calN _{\rm c} \rangle / N_{\rm s}$,
        where $N_{\rm s}$ denotes the number of sites in the lattice. 
    \item The nearest-neighbor connectivity $\gn$, defined by 
      \begin{equation}
        \calE_{\rm n}(\calA) = \sum_{xy \in \{e_{ij}\}} \gamma_{x,y} (\calA)\; , \hspace{10mm} 
        \gn = \langle \calE_{\rm n} \rangle / N_{\rm n} \; , \nonumber
      \end{equation}
	where the sum is on all nearest-neighbor pairs, 
and $N_{\rm n} = N_{\rm e}$ is the total number of nearest-neighbor pairs.
    \item The next-nearest-neighbor connectivity $\gnn$, defined analogously as $\gn$, 
except that the summation on $xy$ involves next-nearest-neighbor pairs, 
and that the denominator $N_{\rm n}$ is replaced by the total number of next-nearest-neighbor
pairs of the lattice. Connectivities at other distances can be defined similarly.
  \end{itemize}
\item {\it Specific-heat-like quantities}: 
  \begin{itemize}
    \item $\Cn =  \left(\langle \calE^2_{\rm n} \rangle - \langle  \calE_{\rm n} \rangle^2 \right) / N_{\rm e} $. 
    \item $\Cnn$, defined analogously as $\Cn$ for the next-nearest neighbors.
  \end{itemize}
\end{enumerate}

\subsection{Finite-size scaling} 
\label{fss}

The analysis of the sampled quantities, obtained by numerical simulation of the
percolation model, is based on FSS predictions. To obtain these predictions,
one first expresses these quantities in terms of the derivatives of the free-energy
density $f=- L^{-d} \ln Z$  of the random-cluster model with respect to the
thermal field $t$, the magnetic field $h$, or the parameter $q$. Then, one applies
the scaling relation for the free-energy density $f(q,t,h,L)$, which is
\begin{equation}
  f(q,t,\,h,\,L)= f_{\rm r}(q,t,\,h)+ L^{-d} f_{\rm s}(q,t L^{y_t},\, hL^{y_h}, 1) \;, 
\label{eq:grand-f}
\end{equation}
where the irrelevant scaling fields have been neglected,
$f_{\rm r}$ denotes the regular part of the free-energy density, and $f_{\rm s}$ is the singular part. 
The thermal scaling field $t$ is approximately proportional to $u-u_{\rm c}$, 
where $u_{\rm c}$ is the critical value of $u$.

Differentiation of the partition sum~(\ref{zrc}) with respect to $u$
at the critical point shows that 
\begin{eqnarray}
 (-u L^d/N_{\rm e}) (\partial f/\partial u) &=& N_{\rm e} ^{-1} 
  \langle \calN_{\rm b} \rangle \equiv \rho_{\rm b} (L)
  = \rho_{{\rm b},0} +  a L^{y_t-d} \;, \label{eq:scaling-rhob} 
\end{eqnarray}
where $\rho_{{\rm b},0}$ represents the bond density in the thermodynamic limit. 
The last equality in Eq.~(\ref{eq:scaling-rhob}) follows from Eq.~(\ref{eq:grand-f}).
In the $q \to 1$ limit, the amplitude $a$ vanishes as $a \approx a_1(q-1)$.

The FSS behavior of the nearest-neighbor connectivity $\gn (L)$ follows from its relation
with $\rho_{\rm b} (L)$. The mapping on the random-cluster model \cite{KF} shows that 
Potts variables in the same cluster are equal, variables in different clusters are
uncorrelated, and that each Potts pair of nearest neighbors is connected by a bond
with probability $p \delta_{\sigma_i \sigma_j}$, where $p \equiv u/(u+1)$. 
The fraction $g_{\rm n}$ of the nearest neighbors belonging to the same cluster thus
contribute a term $p g_{\rm n}$ to the bond density. The remaining fraction $1-g_{\rm n}$
of nearest-neighbor pairs lie across a boundary between two different clusters, and
there is still a probability $1/q$ that the two spins of the pair are equal.
The latter pairs thus contribute a second term $p (1-g_{\rm n})/q$ to the bond density.
Therefore, for integers $q>1$, the bond density is expressed in $\gn$ as
\begin{equation}
\rho_{\rm b}  =p[\gn + (1-\gn)/q] \;.
\label{rhob-gn}
\end{equation}
It follows from Eq.~(\ref{eq:scaling-rhob}) and (\ref{rhob-gn}) that, 
at criticality $p=p_{\rm c}$, $\gn$ is given by 
\begin{equation}
  \gn (L) = \frac{ {q \rho_{{\rm b},0}}/{p_{\rm c}} - 1  } {q-1} 
+ \frac{q a L^{y_t-d} }{(q-1)p_{\rm c}} \; .
\label{eq:scaling-gnL}
\end{equation}
One expects that this expression remains valid for non-integer values of $q$.
We denote the first term in Eq.~(\ref{eq:scaling-gnL}) by $g_{ {\rm n},0}$, and postpone
its evaluation to Sec.~\ref{exact-gn}.
In the limit $q \to 1$, it is sufficient to linearize the amplitude $a$ as
$a \simeq a_1(q-1)$ which yields:
\begin{equation}
  \gn (L) = g_{ {\rm n},0} + g_{{\rm n},1} L^{y_t-d} \;, 
\label{eq:scaling-gn}
\end{equation}
where the amplitude $g_{{\rm n},1}$ takes a nonzero value $a_1/p_{\rm c}$.
The above equation expresses that, in spite of the suppression of 
the critical thermal fluctuations,
$\gn (L)$ does display a singular dependence on $L$. 
Similar FSS behavior is expected for $\gnn(L)$.

For the specific-heat like quantities $\Cn$ and $\Cnn$ at criticality, 
one might simply expect
\begin{equation}
\Cn(L) \sim \Cnn(L) \propto C_0 + c_1 L^{2y_t-d}.
\label{eq:FSS-Cn}
\end{equation}
As numerically demonstrated later, Eq.~(\ref{eq:FSS-Cn}) does not hold 
exactly, namely, a term proportional to $L^{2y_t-d} \ln L$ is present. 
We will explain the logarithmic factor by relating $\Cn$ to 
observables whose two-point functions scale logarithmically for 
$q \rightarrow 1$ \cite{VJS, VJ14}.

\section{Exact values for the connectivity $\gn$ in the thermodynamic limit}
\label{exact-gn}

At criticality Eq.~(\ref{eq:scaling-gnL}) yields, in the thermodynamic limit,
\begin{equation}
  \gnz  = \frac{ {q \rho_{{\rm b},0}}/{p_{\rm c}} - 1  } {q-1} \;.
\label{eq:exact-gn}
\end{equation}
Using this formula, and the known behavior of $\rho_{{\rm b},0}(q)$ and $p_{\rm c}(q)$,
exact values of $\gn$ can be derived.
On the square lattice, the condition of self-duality yields the critical 
parameters $\rho_{{\rm b},0}(q)=1/2$ and $p_{\rm c}=\sqrt{q}/(\sqrt{q}+1)$. 
Thus for general values of $q$, one has
\begin{equation}
  \gnz(q)  = \frac{\sqrt{q}+2}{2(\sqrt{q}+1)} \;,
\label{eq:exact-gnq}
\end{equation}
which yields $\gnz=3/4$ for the bond-percolation problem ($q=1$).

For the triangular lattice one has $\langle \delta_{\sigma_i \sigma_j} \rangle = -E/3K$,
where $E$ is the reduced internal energy. 
The critical value of $K$ as a function of $q$ is given in Ref.~\onlinecite{KJ74}, 
and that of $E$ is given in Ref.~\onlinecite{BTA78}.
At criticality, 
considering $\rho_{{\rm b},0}=p_{\rm c} \langle \delta_{\sigma_i \sigma_j} \rangle$,
the substitution of $K_{\rm c}(q)$ and $E_{\rm c}(q)$ into Eq.~(\ref{eq:exact-gn}) yields 
the function $\gnz^{\rm tri}(q)$ as 
 \begin{equation}
 \gnz^{\rm tri}(q) =  \frac{ -{q E_{\rm c}/3 K_{\rm c} - 1  }} {q-1} \;.
 \label{eq:tri-gnq}
 \end{equation}
For the honeycomb lattice, which is dual to the triangular lattice, the function
$\gnz^{\rm hon}(q)$ can be obtained from its duality relation with $\gnz^{\rm tri}(q)$.    
The relation tells that if there is a (no) bond on an edge of the triangular
lattice, there will be no (a) bond on the dual edge in the honeycomb lattice.
Furthermore, if there is no bond between two nearest-neighbor sites,  then,
if the two sites are connected (disconnected), the dual pair of sites 
will be disconnected (connected).

Taking the $q \rightarrow 1$ limit of Eq.~(\ref{eq:tri-gnq}), 
one can derive (see Appendix~\ref{gnHonDer}) that 
$\gnz^{\rm tri} 
={3(2+7 p_c^{\rm tri})}/{4(5-p_c^{\rm tri})}
=0.714\;274\;133\,\cdots$
for the bond-percolation problem on the triangular lattice, 
and making use of the duality relation, we obtain 
$\gnz^{\rm hon}
= {3 (-2+9p_c^{\rm hon})}/{4(1+4p_c^{\rm hon})}
= 0.804\;735\;202\,\cdots$
for the honeycomb lattice, where $p_c^{\rm tri}=2\sin(\pi/18)$
and $p_c^{\rm hon}=1-2\sin(\pi/18)$ \cite{SE64} are bond-percolation thresholds 
for the two lattices, respectively. 
Noting that $p_c^{\rm tri}$ is the solution to $p^3-3p+1=0$ \cite{SE64},
substituting the relations between $\gnz$ and $p_c^{\rm tri}$ into the cubic equation,
it can be derived that the $\gnz^{\rm tri}$ and $\gnz^{\rm hon}$ are solutions to 
cubic equations 
$64 x^3 -144 x^2 -144x + 153=0$ and $64 x^3 -432 x^2 -720x + 333=0$, respectively. 
These results are similar to those of Ref.~\onlinecite{ZFA97}, 
where the results of Ref.~\onlinecite{BTA78} for the cluster-number densities
on the triangular and honeycomb lattices are written in terms of $p_c$ of 
the two lattices, and identified as solutions to cubic equations. 

\section{Numerical Results}
\label{Num-results}
To confirm the exact values of $\gn$, and to explore the FSS properties,
we simulated the bond percolation models on the square, triangular, honeycomb, 
and simple-cubic lattices. The results are presented in the following subsections.

\subsection{Finite-size analysis for the square lattice}
The Monte Carlo simulations of the bond-percolation model on $L \times L$ square
lattices with periodic boundary conditions follow the standard procedure:
each edge is randomly occupied by a bond with the critical probability 
$p=p_{\rm c}=1/2$, and the resulting bond configuration is then decomposed 
in percolation clusters.
Quantities are sampled after every sweep. The simulations used $22$ sizes 
in range $4 \leq L \leq 8000$, with numbers of samples around 
$100$ million for $L \le 120$, $80$ million for $160 \le L \le 480$,
$50$ million $L=800$, $25$ million for $L=1600$, $10$ million for
$L=4000$ and $2.5$ million for $L=8000$. 
Roughly $22$ months of computer time were used.

\subsubsection{Connectivities $\gn$ and $\gnn$}
\label{conn_analy}
We fitted our Monte Carlo data for $\gn$ by the formula 
\begin{equation}
\gn=\gnz + L^{y_t-d} (c_1 + c_2 L^{y_i}),
\label{fit_gn}
\end{equation}
with $y_t=3/4$, $y_i=-2$ and $d=2$.
Extrapolations are conducted by successively  removing the first few
small-size data points, while using the guidance of the $\chi^2$ criterion.
The results are $\gnz =0.749\; 999\; 99(13)$,
$c_1=0.277\; 6(3)$ and $c_2=-0.15(13)$, with $L_{\rm min}=16$.
These error margins in the numerical results are quoted as two standard deviations, 
and include statistical errors only. 
The $\gnz$ value is in perfect consistency with the assumption of 
the continuity of $\gn(L = \infty)$ in Eq.~(\ref{eq:scaling-gnL}) as a 
function of $q$, used to derive $\gn(L=\infty)=3/4$ in the limit $q \to 1$.

The fit of the $\gnn$ data, using the same scaling formula, Eq.~(\ref{fit_gn}),
yielded $\gnnz=0.687\;500\;0 (2)$, $c'_1=0.416\; 5 (4)$ and $c'_2=-0.31 (17)$, 
with $L_{\rm min}=16$. The precision of $\gnnz$ supports the conjecture that
$\gnn (L=\infty) =11/16$ holds exactly. This reproduces a conjecture \cite{MN}
for correlations in the completely packed $O(1)$ loop model, which was
based on exact results for correlations on $L \times \infty$
cylinders for several finite $L$ values. This $O(1)$ loop model can be mapped on
the square-lattice percolation model on a cylinder, but with the axis of
the cylinder along a diagonal direction of the square lattice.
In Appendix~\ref{appMapping} we describe the relation between $\gnn$ in 
the percolation model
and the probability that two consecutive points lie on the same loop of 
the completely packed $O(1)$ loop model.

The Monte Carlo data for $\gn$ and $\gnn$ are presented in Table~\ref{fludat}.
We also performed some transfer-matrix calculations of these two quantities 
in $L \times \infty$ bond-percolation systems. These show that the connectivities
converge very quickly to their infinite-system values $3/4$ and $11/16$
as $L$ increases. The finite-size results for $\gn$ and $\gnn$ are
obtained as fractional numbers, which reflects the interesting algebraic properties
already observed in the related context of the completely packed $O(1)$ loop
model \cite{MN}.  These results are presented in Appendix~\ref{TMcal}.

\subsubsection{Numerical evidence of a logarithmic factor in the 
scaling behavior of $\Cn$ and  $\Cnn$}
\label{logCn}

For the quantity $\Cn$, which describes the amplitude of the fluctuations in $\gn$, 
we tried several fits according to
\begin{equation}
\Cn=\Cnz + L^{\psi}(d_1+d_2 L^{y_i}+\cdots).
\label{fit_Cn}
\end{equation}
The results suggest that $\Cnz \approx 4.22$ and $\psi \approx -0.358$.
For example, a fit to the data by $\Cnz+C_{\rm n,1} L^{\psi}$ yielded
$\Cnz=4.20(4)$, $C_{\rm n,1}=-6.2(6)$, $\psi=-0.368(13)$, with 
$L_{\rm min}=320$ for the cutoff at small system sizes. 
However, some caution concerning the result $\psi \approx -0.358$ for
the leading finite-size exponent in $\Cn$ seems justified. 
Apart from the fact that the exponent $-0.358$  cannot be expressed as a suitable
combination of the renormalization exponents and the space dimensionality $d$,
acceptable values of $\chi^2$ could only be obtained for unusually large
$L_{\rm min}$.

\begin{figure}
\centering
\includegraphics[scale=]{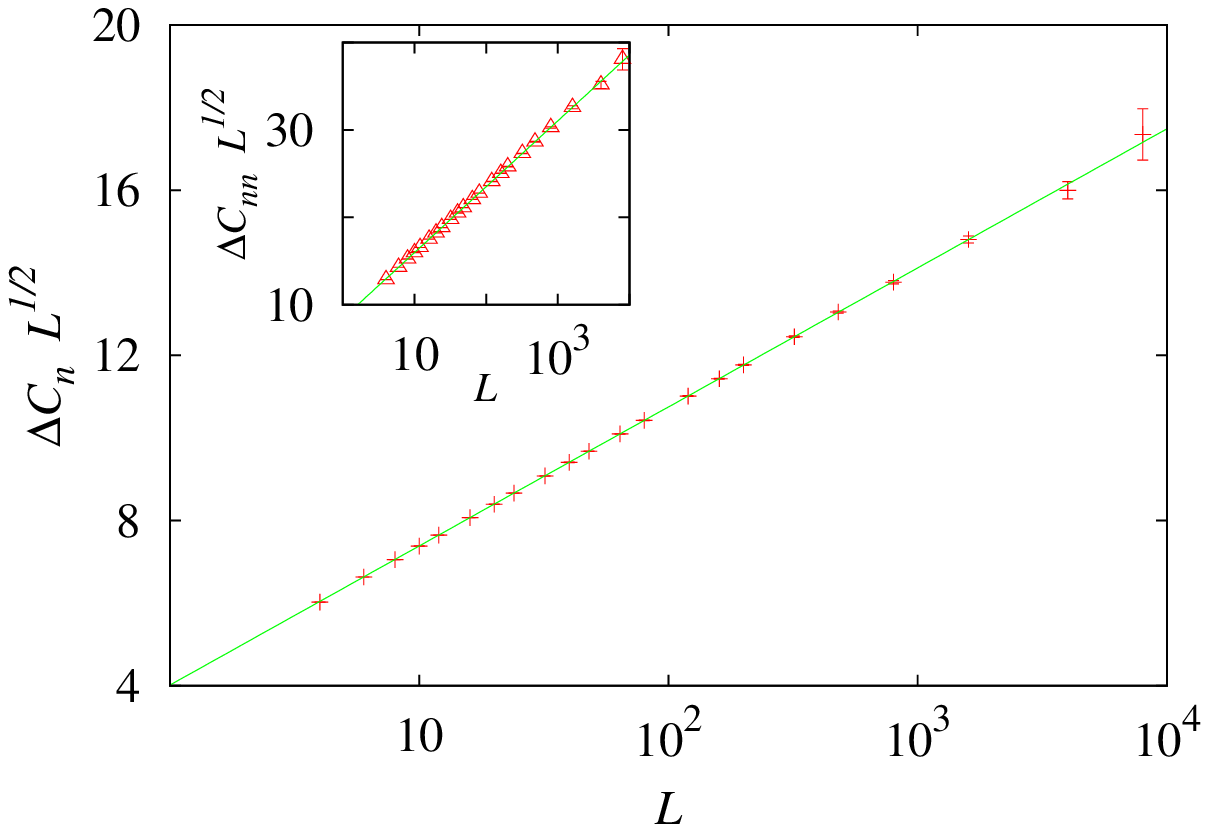}
\caption{The quantities $\Delta C_x L^{1/2} 
= (C_x (\infty) - C_x (L)) L^{1/2}$, where $x$ represents `n' and `nn' (inset),
{\em versus} system size $L$ on a logarithmic scale.
The quantities $\Cn(L)$, $\Cnn(L)$ are the amplitudes of the fluctuations in 
the nearest- and the next-nearest-neighbor connectivities respectively,
for bond percolation on $L \times L$ square lattices with
periodic boundary conditions.
We use $\Cn (\infty) = 4.169\;8$ and $\Cnn (\infty) = 8.206$, 
as determined by our fits.
The figures clearly indicate the presence of a logarithmic factor in the
leading scaling term of $\Cn$ and $\Cnn$. The lines are added for clarity.}
\label{grand_con}
\end{figure}

Since, as will be argued in Sec.~\ref{originLog}, a multiplicative
logarithmic factor may occur in the singular behavior of $\Cn$, 
we also applied fits according
to $\Cn=\Cnz + d_1 L^{y_1} \ln L + d_2 L^{y_2}$.
For $y_1=y_2$ this reduces to 
$\Cn = \Cnz + d_1 L^{y_1} \ln L/L_0 + \cdots $, 
with $d_2=-d_1 \ln L_0$.
With fixed $y_1=y_2=2y_t-d=-1/2$, the fit 
led to $\Cnz=4.169\;8(12),~d_1=-1.462(4),~d_2=-4.018(5)$, 
with $L_{\rm min}=8$.  
Other fits with $y_1$ or $y_2$ as free parameters yielded consistent results.
One observes that the fits including a logarithm use fewer parameters
and/or a smaller cutoff $L_{\rm min}$. This indicates 
that a multiplicative logarithmic factor indeed appears in the scaling of $\Cn$.
We present our data for this quantity in Table~\ref{fludat}. The existence of
the logarithmic factor in these data is illustrated in Fig.~\ref{grand_con}.

\begin{table}
\begin{center}
    \begin{tabular}{|c|c|c|c|c|c|c|}
    \hline
       $L$& 4 & 8 & 16 & 32 & 64 & 120 \\ 
    \hline
        $\gn$ &  $0.797\;65(2)$ & $0.770\;51(1)$ & $0.758\;66(1)$ &
		 $0.753\;643(4)$ & $0.751\;536(2)$ & $0.750\;698(1)$ \\ 
	$\gnn$ & $0.755\;67(3)$ & $0.718\;11(2)$ & $0.700\;48(1)$ & 
		 $0.692\;964(5)$ & $0.689\;804(3)$ & $0.688\;547(2)$ \\
        $\Cn$ & 
                  $1.156\;8(2)$ & $1.674\;6(2)$ & $2.152\;2(3)$ & 
                  $2.564\;3(4)$ &  $2.907\;8(4)$ & $3.164\;2(5)$ \\
        $\Cnn$ & 
                  $1.763\;9(2)$ & $2.820\;3(4)$ & $3.831\;6(5)$ & 
		  $4.711\;8(6)$ & $5.451\;5(7)$ & $6.0063(8)$ \\
    \hline
    \hline
       $L$& 200 & 480 & 800 & 1600 & 4000 & 8000 \\ 
    \hline
        $\gn$ & 
		$0.750\;368(1)$ & $0.750\;123\;4(3)$ & $0.750\;065\;4(3)$ & 
		$0.750\;027\;4(2)$ & $0.750\;008\;8(1)$ & $0.750\;003\;6(1)$ \\
        $\gnn$ &
		$0.688\;052(1)$ & $0.687\;685\;3(5)$ & $0.687\;598\;1(4)$ &
		$0.687\;541\;1(3)$ & $0.687\;513\;2(2)$ & $0.687\;505\;3(2)$ \\
        $\Cn$ & 
                  $3.337\;5(5)$ & $3.574\;2(6)$ & $3.683\;0(7)$ &
                  $3.799\;6(11)$ & $3.917(2)$ & $3.976(4)$ \\
        $\Cnn$ & 
                  $6.382\;4(10)$ & $6.899\;2(11)$ & $7.136(2)$ &
		  $7.391(2)$ & $7.650(3)$ & $7.780(7)$ \\
    \hline
    \end{tabular}
\end{center}
\caption{Data for the nearest- and next-nearest- neighbor connectivities
and the amplitudes of their fluctuations for  the bond-percolation model. These data apply
to $L \times L$ systems on square lattices with periodic boundary conditions. 
The quoted error bar corresponds to one standard deviation.}
\label{fludat}
\end{table}

For the quantity $\Cnn$, which represents the amplitude of the fluctuations in $\gnn$,
a fit by $C_{\rm nn,0}+C_{\rm nn,1} L^{\psi}$ led to $C_{\rm nn,0}=8.30(3)$, 
$C_{\rm nn,1}=-12.8(3)$ and $\psi=-0.358(3)$, 
necessarily with a cutoff at a large size $L_{\rm min}=200$. 
These results tell that the FSS behavior of $\Cnn$ on the square lattice is 
similar to that of $\Cn$.
A fit to the data by $C_{\rm nn,0} + d'_1 L^{-1/2} \ln L + d'_2 L^{-1/2}$
yielded $C_{\rm nn,0}=8.206(2)$, $d'_1=-3.271(7)$ and $d'_2=-8.43(1)$, 
with $L_{\rm min}=8$. Other fits with either/both of the exponents 
as free fitting parameters also yielded results consistent with those for $\Cn$.
Thus, also the results for $\Cnn$ 
indicate the existence of a logarithmic factor. Data for $\Cnn$ 
are also presented in Table~\ref{fludat} and plotted in Fig.~\ref{grand_con}.

\subsection{Finite-size analysis for other lattices}
\label{extension}

\subsubsection{Triangular lattice with periodic boundary conditions}
We simulated the bond-percolation problem on the triangular lattice 
at the percolation threshold $p^{\rm tri}_{\rm c}=2 \sin (\pi/18)$ \cite{SE64}.
Rhombus-shaped lattices were used, with periodic boundary conditions 
applied along edges of the rhombus. 
We used lattices with $L^2$ sites, with $7$ different values of 
the linear size $L$ in the range between $18$ and $1152$. 
The number of samples was $100$ million for each size.
Fits of the $\gn$ data by $g^{\rm tri}_{n,0} + c_1 L^{y_1}$ yielded 
$g^{\rm tri}_{n,0}=0.714\;273\;9(8)$ and $y_1=-1.249(7)$. 
The value $y_1$ agrees well with $y_t-d=-5/4$. 
Fits of the data by Eq.~(\ref{fit_gn}) led to 
$\gnz^{\rm tri}=0.714\;273\;9(3)$, 
which is in good agreement with the theoretical prediction 
$0.714\; 274\; 133\; \cdots$ in Sec.~\ref{exact-gn}. 
For $\gnn$, fits of the data by Eq.~(\ref{fit_gn}) yielded
$\gnnz^{\rm tri}=0.637\;428\;6(5)$.

A fit of the $\Cn$ data by $C^{\rm tri}_{n,0} + c_1 L^{\psi}$ led to 
$C^{\rm tri}_{n,0}=7.24(3)$, $c_1=-10.0(3)$ and $\psi=-0.351(8)$, with $L_{\rm min}=144$.
The value $-0.351$ of the exponent is quite different from $2y_t-d=-1/2$.
When including a logarithmic factor, a fit of the $\Cn$ data 
by $C^{\rm tri}_{n,0} + c_1 L^{-1/2} + c_2 L^{-1/2} \ln L$ yielded 
$C^{\rm tri}_{\rm n,0}=7.140(3)$, $c_1=-6.99(3)$ and $c_2=-2.54(2)$, with $L_{\rm min}=18$.  
For $\Cnn$, the first fit led to $\psi=-0.343(7)$, with $L_{\rm min}=144$, 
and the second fit yielded $C^{\rm tri}_{\rm nn,0}=17.595(11)$, $c_1=-18.5(2)$ 
and $c_2=-7.67(7)$, with $L_{\rm min}=36$.

The above results tell that the FSS behavior of the connectivities and their
fluctuations on the triangular lattice is similar to that on the square lattice. 

\subsubsection{Honeycomb lattice with periodic boundary conditions}

We also simulated the bond-percolation problem on the honeycomb lattice 
at the percolation threshold $p^{\rm hon}_{\rm c}=1-2 \sin (\pi/18)$. 
Rhombus-shaped lattices were used, with periodic boundary conditions
applied along edges of the rhombus.
We used lattices with $L^2/2$ sites, with $8$ different values of 
the linear size $L$ in the range between $8$ and $1024$. 
The number of samples was $100$ million for each size.
Fits of the $\gn$ data by $g^{\rm hon}_{n,0} + c_1 L^{y_1}$ yielded 
$g^{\rm hon}_{n,0}=0.804\;735\;3(2)$ and $y_1=-1.250(1)$. 
The value $y_1$ agrees well with $y_t-d=-5/4$, and the numerical value of $\gnz^{\rm hon}$
is in good agreement with the theoretical prediction 
$0.804\; 735\; 202\; \cdots$ in Sec.~\ref{exact-gn}. 

A fit of the $\Cn$ data by $C^{\rm hon}_{n,0} + c_1 L^{\psi}$ led to 
$C^{\rm hon}_{n,0}=2.367(6)$, $c_1=-2.94(4)$ and $\psi=-0.351(5)$, with $L_{\rm min}=128$.
The value $-0.351$ of the exponent is quite different from $2y_t-d=-1/2$.
When including a logarithmic factor, a fit of the $\Cn$ data 
by $C^{\rm hon}_{n,0} + c_1 L^{-1/2} + c_2 L^{-1/2} \ln L$ yielded 
$C^{\rm hon}_{n,0}=2.332\;8(6)$, $c_1=-2.170(4)$ and $c_2=-0.722(2)$, with $L_{\rm min}=16$.
Thus, as expected,  the FSS behavior on the honeycomb lattice is similar to that 
on the square and triangular lattices. It indicates that the logarithmic factor 
is  a universal property of two-dimensional lattices.

\subsubsection{The three-dimensional cubic lattice}

The bond-percolation model on three-dimensional $L^3$
simple-cubic lattices with periodic boundary conditions was investigated. 
The simulations were done at $11$ different sizes $4 \leq L \leq 256$,
at a bond-occupation probability $p=p^{\rm cub}_{\rm c}=0.248\;811\;8$ \cite{WZZGD}.
The number of samples was over $100$ million for $L\le 64$, and around $10$ million
for $L \ge 128$.

A fit of the $\gn$ data by $\gnz^{\rm cub} + c_1 L^{y_1}$ led to 
$\gnz^{\rm cub}=0.359\;404\;4(3)$ and $y_1=-1.857\;3(14)$,  
with $L_{\rm min}=24$.  
The $y_1$ value is consistent with $y^{\rm (3)}_t-d \approx -1.859$, as it follows
from the $d=3$ literature value of the thermal exponent, namely
 $y^{\rm (3)}_t = 1.141\;0(15)$ \cite{WZZGD}.

\begin{figure}
\centering
\includegraphics[scale=]{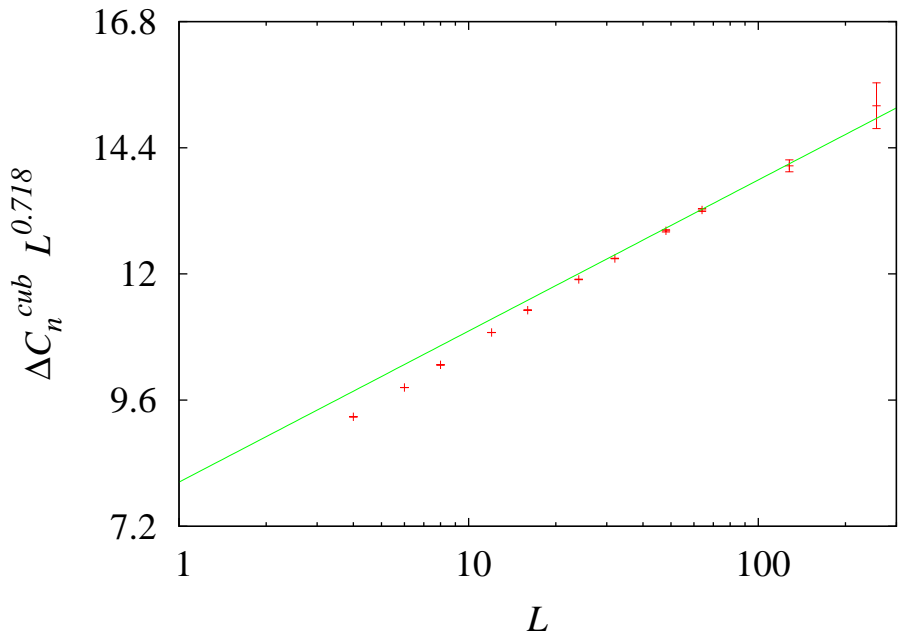}
\caption{The quantity $\Delta \Cn^{\rm cub} L^{0.718} 
= (\Cn^{\rm cub} (\infty) - \Cn^{\rm cub} (L)) L^{0.718}$ 
{\em versus} system size $L$ on a logarithmic scale.  $\Cn^{\rm cub}(L)$ represents 
the amplitude of the fluctuations in the nearest-neighbor connectivities 
for percolation on a $L^3$ simple-cubic lattice with periodic boundary
conditions.  We use $\Cn^{\rm cub} (\infty) = 6.566$ as obtained from the fit.
The line is added for clarity. Its nonzero slope expresses the presence of a
logarithmic factor in the leading finite-size dependence of $\Cn^{\rm cub}$.
Deviations at small $L$ values are attributed to finite-size correction terms.  }
\label{fig-cubCn}
\end{figure}

A fit of the $\Cn$ data by $\Cnz^{\rm cub} + d_1 L^{\psi}$ yielded
$\Cnz^{cub}=6.57(2)$ and $\psi=-0.602(16)$, with $L_{\rm min}=32$. 
Including a correction term with exponent $y^{\rm (3)}_i=-1.2$ \cite{WZZGD},
another fit of the data by $\Cnz^{\rm cub} + L^{\psi} (d_1+d_2 L^{y^{\rm (3)}_i})$ led to 
$\Cnz^{\rm cub}=6.57(2)$ and $\psi=-0.62(2)$, with $L_{\rm min}=16$. 
These values of $\psi$ are different from $2y^{\rm (3)}_t-d \approx -0.718$.
Instead, a fit by $\Cnz^{\rm cub} + L^{-0.718} ( d_1 + d_2 \ln L + d_3 L^{y^{\rm (3)}_i})$
led to $\Cnz^{\rm cub}=6.556(16)$, $d_1=-8.0(7)$ and $d_2=-1.25(22)$,
with $L_{\rm min}=16$.
These fit results support the appearance of a multiplicative logarithmic 
factor in the FSS behavior of $\Cn$, which is also shown in Fig.~\ref{fig-cubCn}.

\subsubsection{The square lattice with open boundaries}

We also performed bond-percolation simulations at a bond-occupation probability
$p=1/2$, using a square $L \times L$ geometry, with periodic boundary conditions in one
direction and open boundary conditions in the other direction. 
We took 11 system sizes from $L=8$ to $L=256$, and a number of 
$100$ million independent percolation configurations for each size,
in order to sample the nearest- and next-nearest-neighbor connectivities
$\gn^{\rm sur}$ and $\gnn^{\rm sur}$ on the open boundaries.
Note that a pair of next-nearest neighbors on the boundary is
separated by a distance of $2$ lattice units, instead of $\sqrt{2}$ as in the bulk.

Fits of the $\gn^{\rm sur}$ data by $\gnz^{\rm sur}+c_1 L^{y_1}$ yielded
$\gnz^{\rm sur}=0.625\;000\;1(12)$ and $y_1=-1.999(6)$;
and fits of the $\gnn^{\rm sur}$ data 
led to $\gnnz^{\rm sur}=0.449\;789(2)$, $c_1=2.22(5)$ and $y_1=-2.005(7)$.
On a free boundary, the scaling dimension of the energy operator $\epsilon$ should
be replaced by $\Delta_{\epsilon}=2$ \cite{Cardy84} ($y_{\epsilon}=d-\Delta_{\epsilon}=0$). 
The numerical results for $\gn^{\rm sur}$ and $\gnn^{\rm sur}$ agree very well 
with $y_1=y_{\epsilon}-d=-\Delta_{\epsilon}$. 
The surface connectivities converge more quickly with the size of the system 
than the bulk ones, possibly because surface clusters are smaller or 
their correlations fall off faster, so that they are not as strongly affected 
by the finite size of the system.

The result for the surface connectivity strongly suggests that $\gnz^{\rm sur}=5/8$
holds exactly.  It applies to a system with a bond probability $p=1/2$ on the
open boundary. When we erase those bonds, the limiting probability that
two nearest-neighboring sites on the boundary are connected decreases to
$\gnz^{\rm sur'}=5/32$, as can be easily checked by adding a row of $p=1/2$
bonds perpendicular to the boundary. Next, we may merge two half-infinite systems,
one with, and one without boundary bonds, thus reconstructing the infinite system.
The combined probability that the two nearest-neighboring sites are now connected
by some path {\em within either system} is $5/8+ 3/8\times 5/32=175/256$, slightly
smaller than the bulk value $3/4$. It thus appears that there is a probability
$3/4-175/256=17/256$ that connections
between the two neighboring sites exist only via paths entering both half systems.

\section{Origin of the logarithmic factor in the finite-size scaling}
\label{originLog}
We show that the quantity $\Cn$ relates to connectivities of four points
$\{\vec{x}_1, \vec{y}_1, \vec{x}_2, \vec{y}_2\}$, 
in which $\vec{x}_1$ and $\vec{x}_2$ are two sites separated by a distance $r$, 
and $\vec{y}_1$, $\vec{y}_2$ denote a nearest neighbor of $\vec{x}_1$ 
and $\vec{x}_2$, respectively. In Ref.~\onlinecite{VJS} a logarithmic term was
derived in the FSS of these connectivities, in
the limit $q \rightarrow 1$. It was obtained  from the mixing of 
the energy operator with the operator that connects two random clusters.
These two operators become degenerate at $q=1$, with  the same scaling
dimension $5/4$ in two dimensions.

Following the notation of Ref.~\onlinecite{VJS}, we define $P_0(r)$ as the 
probability that the sites $\{\vec{x}_1, \vec{y}_1, \vec{x}_2, \vec{y}_2\}$ belong to four different
percolation clusters; $P_1(r)$ as the probability that
$\{\vec{x}_1, \vec{y}_1, \vec{x}_2, \vec{y}_2\}$
belong to three different clusters, of which one cluster connects one of
$\{\vec{x}_1, \vec{y}_1\}$ to one of $\{\vec{x}_2, \vec{y}_2\}$; and $P_2(r)$ as the probability that
the four points belong to two different clusters, each of which 
contains one point of $\{\vec{x}_1, \vec{y}_1\}$ and one point of $\{\vec{x}_2, \vec{y}_2\}$. 
The probability that the pair $\{\vec{x}_1, \vec{y}_1\}$ is unconnected, while  the pair
$\{\vec{x}_2, \vec{y}_2\}$ is simultaneously  unconnected, is equal to
\begin{eqnarray}
P_0(r)+P_1(r)+P_2(r)
&=& \langle (1-\gamma_{x_1 y_1}) (1-\gamma_{x_2 y_2}) \rangle \nonumber\\ 
&=& 1-2 \gn +
     \langle \gamma_{x_1 y_1} \gamma_{x_2 y_2} \rangle \;,
\label{sumP}
\end{eqnarray}
(for convenience, we omit the arrow symbol over the site coordinates here and below).\\
Next, we express the quantity $\Cn$ as 
\begin{eqnarray}
\Cn &=& N_{\rm e}^{-1}   \left(\langle \calE^2_{\rm n} \rangle - \langle  \calE_{\rm n} \rangle^2 \right) \nonumber \\
&=& N_{\rm e}^{-1}   \sum_{x_1 y_1,\; x_2 y_2}
\left( \langle \gamma_{x_1 y_1} \gamma_{x_2 y_2} \rangle  
- \langle \gamma_{x_1 y_1} \rangle \langle \gamma_{x_2 y_2} \rangle \right)  \nonumber \\
&=& \left( N_{\rm e}^{-1}   \sum_{x_1 y_1 \neq x_2 y_2} \langle \gamma_{x_1 y_1} \gamma_{x_2 y_2} \rangle \right) + \langle \gamma_{x_1 y_1} \rangle 
- N_{\rm e} \langle \gamma_{x_1 y_1} \rangle^2 \nonumber \\ 
&=& \left( N_{\rm e}^{-1}   \sum_{x_1 y_1 \neq x_2 y_2} \langle \gamma_{x_1 y_1} \gamma_{x_2 y_2} \rangle \right) + \gn
- N_{\rm e} \gn^2 \;.
\label{logTermCn}
\end{eqnarray}
The FSS singularity of $\Cn$ resides in the first term in the
last line of Eq.~(\ref{logTermCn}), in particular in the dependence of
$\langle \gamma_{x_1 y_1} \gamma_{x_2 y_2} \rangle$ on the distance $r$ between
$(x_1 y_1)$ and $(x_2 y_2)$.
Using Eq.~(\ref{sumP}), and considering that $1-\gn$ equals the probability
that two neighboring points belong to different clusters, one derives
\begin{eqnarray}
\langle \gamma_{x_1 y_1} \gamma_{x_2 y_2} \rangle _{r}
&=& \left( P_0(r)+P_1(r)+P_2(r)-1+2 \gn  \right)  \nonumber \\
&=& \left( P_0(r)+P_1(r)+P_2(r)-(1-\gn)^2 \right) + \gn ^2 \;.
\label{logTerm}
\end{eqnarray}
According to Ref.~\onlinecite{VJS}, it behaves as $\langle \gamma_{x_1 y_1}
\gamma_{x_2 y_2} \rangle_{r} \simeq \gn^2 +(a+b\ln r)r^{-2 \Delta}$ in two dimensions,
where $\Delta=5/4$ is the common scaling dimension of the two degenerate operators.
The scaling behavior of the sum is therefore 
\begin{displaymath}
N_{\rm e}^{-1}\sum_{x_1 y_1 \neq x_2 y_2}\langle \gamma_{x_1 y_1} \gamma_{x_2 y_2}\rangle
\approx(N_{\rm e}-1)\gn^2+\int_1^{L/2}2\pi rdr(a+b \ln r) r^{-5/2}
\end{displaymath}
\begin{equation}
=(N_{\rm e}-1)\gn^2+\left(A+B\ln L\right)L^{-1/2}\;,
\end{equation}
where $A$ and $B$ are non-universal constants.
Substituting the above result in Eq.~(\ref{logTermCn}), one gets
\begin{eqnarray}
\Cn(L) &=& \Cn(\infty) +d_1 L^{-1/2} \ln L  + d_2 L^{-1/2} + \cdots \;.
\label{resultCn}
\end{eqnarray}
This explains the multiplicative logarithmic factor in the singular part of $\Cn$.

Eq.~(\ref{resultCn}) still contains a contribution due to $\gn$, which,
as noted in Sec.~\ref{fss}, satisfies $\gn=g_{n,0}+c_1 L^{y_t-d}+o(L^{y_t-d})$.
The terms in Eq.~(\ref{resultCn}) originating from $\gn$ thus
contribute a constant contained in $\Cn(\infty)$, and the omitted terms
include one proportional to  $L^{y_t-d}$, etc.
This conclusion is consistent with the numerical results in the previous section.

Similar arguments apply in the case of $\Cnn$.
The above analysis is not restricted to the two-dimensional case. 
Indeed, a similar relation between $\Cn$ and the four-point connectivities 
holds for $d>2$; and it is expected that the energy operator and the operator which connects
two random clusters become degenerate also in higher dimensions \cite{VJS, VJ14, Con82}. 
Thus we expect a logarithmic factor also for $d>2$ in the FSS behavior of $\Cn$, 
which is supported by our numerical results for 
the three-dimensional cubic lattice in the previous section.

\section{Discussion}
\label{discus}
As already clear from the work of Mitra et al.~\cite{MN}, critical
connectivities in the percolation model display remarkable algebraic 
properties. Completely in line with these findings are the results for the
exact eigenvectors in Appendix~\ref{TMcal}, the exact value 
$\gn=3/4$, and the conjectured exact values $\gnn=11/16$ and $\gn^{\rm sur}=5/8$ 
for the square-lattice model.  We also derived, from the existing results for the
Potts model, the exact values of $\gn$ on the triangular and honeycomb lattices. 
Results from Monte Carlo simulations agree very well with these exact or conjectured values. 
In addition, we numerically determined some other neighboring connectivities. 
Our results for critical short-range connectivities in the thermodynamic limit are summarized 
in Tables~\ref{summary-conn-A} and \ref{summary-conn-B}.
For the RC model with $q\ne1$, the critical nearest-neighbor connectivity can be 
obtained from Eq.~(\ref{eq:exact-gnq}) and (\ref{eq:tri-gnq}),  for the square
and triangular lattices, respectively; and for the honeycomb lattice, the nearest-neighbor
connectivity can be obtained from its duality relation with that of the triangular lattice.

\begin{table}
\begin{center}
    \begin{tabular}{|c|l|l|}
    \hline
        lattice	& $\gn$ 	& $\gnn$ \\ 
    \hline
	square	& $0.749\;999\;99(13)$ & $0.687\;500\;0(2)$ \\ 
		& $3/4$		& $11/16 ^*$ (Ref.\onlinecite{MN}) \\
	triangular & $0.714\;273\;9(3)$ & $0.637\;428\;6(5)$ \\
		   & $3(2+7p_c^{\rm tri})/4(5-p_c^{\rm tri})$ &  \\
	honeycomb  & $0.804\;735\;3(2)$ &  \\ 
		   & $3(-2+9p_c^{\rm hon})/4(1+4p_c^{\rm hon})$ &  \\  
	square (surface)& $0.625\;000\;1(12)$ & $0.449\;789(2)$ \\
		        & $5/8 ^*$ & \\
        simple  cubic   & $0.359\;404\;4(3)$ &  \\	
    \hline
    \end{tabular}
\end{center}
\caption{Critical nearest- and next-nearest-neighbor connectivities 
for bond-percolation in different lattices.
For each lattice, the first line shows the numerical result(s), 
and the second line (if applicable) presents the exact or 
conjectured (labeled by `*') value(s). Periodic boundary conditions are used, 
except for the `square surface,' where the connectivities are measured 
on free one-dimensional surfaces of the square lattice.}
\label{summary-conn-A}
\end{table}

\begin{table}
\begin{center}
   \begin{tabular}{|c||c|l||c|l|}
   \hline
             &$x,y$& square             &$x,y$& triangular \\ 
   \hline
$g_{\rm n3}$ & $2,0$ & $0.649\;577\;2(6)$ & $2,0$ & $0.619\;666\;5(6)$ \\ 
$g_{\rm n4}$ & $2,1$ & $0.629\;978\;1(7)$ & $5/2,\sqrt{3}/2$ & $0.584\;475\;2(7)$ \\ 
$g_{\rm n5}$ & $2,2$ & $0.599\;838\;6(8)$ & $3,0$ & $0.569\;576\;8(8)$ \\ 
$g_{\rm n6}$ & $3,0$ & $0.595\;566\;4(8)$ & $3,\sqrt{3}$ & $0.552\;726\;4(8)$ \\ 
$g_{\rm n7}$ & $3,1$ & $0.587\;653\;5(8)$ & $7/2,\sqrt{3}/2$ & $0.548\;185\;7(10)$ \\ 
$g_{\rm n8}$ & $3,2$ & $0.571\;119\;2(9)$ & $4,0$ & $0.536\;507\;7(9)$ \\ 
$g_{\rm n9}$ & $4,0$ & $0.560\;360\;4(12)$& $4,\sqrt{3}$ & $0.526\;981\;8(10)$ \\ 
    \hline
    \end{tabular}
\end{center}
\caption{Critical third to 9th nearest-neighbor connectivities for bond-percolation 
in the periodic square and triangular lattices. The displacement vectors of the
connectivities are listed in Cartesian coordinates under $x,y$. }
\label{summary-conn-B}
\end{table}

In this work, for the percolation model, 
we also investigated the FSS behavior of the short-range connectivities and their fluctuations.
As far as we know, the fluctuation amplitudes $\Cn$ and $\Cnn$ have not yet
been studied before. While $\gn$ and $\gnn$ are energy-like
quantities with leading FSS term proportional to $L^{y_t-d}$,
so that their fluctuations may be expected to have a leading
scaling exponent $2y_t-d$, the analysis using a simple power of
the system size yields a numerical exponent that is very different from $2y_t-d$.
This numerical exponent does not seem to fit a combination of the dimensionality 
and the thermal scaling dimension of the percolation problem.  
However, as described above, satisfactory fits (as judged from the $\chi ^2$
criterion) are obtained by including a logarithmic factor, for
$\Cn$ as well as for $\Cnn$. 
These results support that the fluctuations in the neighboring connectivities scale as 
$C - C_0 \simeq L^{2y_t-d} \ln (L/L_0)$, where $C_0$ is the value of the fluctuations
in the thermodynamic limit, and $L_0$ is a non-universal factor.   
We have thus shown the existence of a class of observables in critical
percolation with logarithmic factors in their scaling behavior,
which are closely related to recently identified four-point 
connectivities which scale logarithmically in critical percolation \cite{VJS, VJ14}.
The origin of the logarithmic factor is different from a mechanism 
which introduces logarithmic factors through the $q$-dependence of the
critical exponents in some critical singularities in percolation \cite{FDB}.
From another point of view, the observed FSS behavior may be used to determine 
the critical exponent $y_t$ in $d>2$ dimensions, where an exact value of $y_t$ 
may not be available. For example, from our results of $\gn$ for the simple-cubic
lattice, the value of $y_t$ for $d=3$ is obtained as $1.142\;7(14)$, 
which is comparable with a latest result $1.141\;0(15)$~\cite{WZZGD},
and consistent with the value $8/7$ conjectured by Ziff and Stell (see \cite{LZ98}).

\acknowledgments
This work is supported by the National Natural Science Foundation of China 
under Grant No. 11275185, and the Chinese Academy of
Sciences. Y. J.\ Deng acknowledges the Ministry of Education (China) for 
the Specialized Research Fund for the Doctoral Program 
of Higher Education under Grant No. 20113402110040 and the Fundamental Research 
Funds for the Central Universities under Grant No. 2340000034. 
The authors thank B.\ Nienhuis for suggesting the transfer-matrix approach to
determine the short-range connectivities, T.\ Garoni for helping on the calculation of 
the exact nearest-neighbor connectivity on the triangular lattice, 
and J. L. Jacobsen for drawing our attention to Ref.~\onlinecite{VJ14}. 
They also thank J. F.\ Wang and Z. Z. \ Zhou for sharing their data 
for bond percolation on the cubic lattice.  One of us (HB) thanks the Department
of Modern Physics of the University of Science and Technology of China in Hefei
for hospitality extended to him.

\appendix
\counterwithin{figure}{section}

\section{Specific-heat behavior in the limit $q \rightarrow 1$}
\label{percsph}
\begin{figure}
\centering
\includegraphics[scale=1.0]{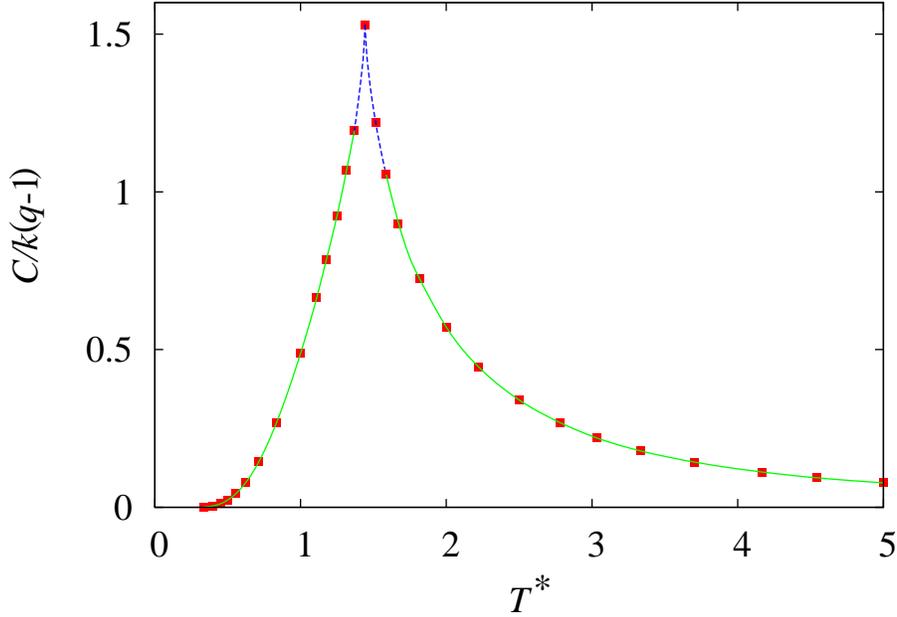}
\caption{(Color online). Dimensionless specific heat $C/k$ of the $q$-state Potts model
on the square lattice, divided by $q-1$, {\em versus} reduced temperature
$T^{*}\equiv kT/J=1/K$, in the percolation limit $q \to 1$. The Potts specific heat
vanishes near $q=1$ as $q-1$, so that the normalization factor
$1/(q-1)$ compensates the vanishing specific-heat amplitude.
The data points (full squares) were obtained by extrapolations of
finite-size data to the thermodynamic limit.  Estimated error bars
do not exceed the symbol sizes. The curves are added for visual aid only.
The critical singularity is governed by a specific-heat exponent $\alpha=2-2/y_t=-2/3$. 
The dashed parts (blue) of the curves display the power-law behavior with this exponent.
In contrast with the Ising model ($q=2$), the specific heat remains finite at the
critical point; this illustrates the ``nonuniversal'' behaviour of the Potts model
when $q$ is varied.
}
\label{fig:pcplt}
\end{figure}

The fluctuations in the energy-like quantities have been used to obtain
specific-heat-like quantities, but thus far we have not considered the actual
Potts model specific heat $C$ per site, which can be expressed as the dimensionless
quantity $C/k \equiv K^2\partial^2 f(K,q)/\partial^2 K$, where $k$ is the  Boltzmann
constant and $f(K,q)\equiv {N_{\rm s}}^{-1} \ln Z(K,q)$ is the reduced free-energy density. 
While the specific heat of the random-cluster model vanishes at $q=1$,
one may still ask the question how it behaves in the limit $q \to 1$. 
From Eq.~(\ref{Potts}) one reads that the energy change associated with
the ordering of the Potts model, i.e., the integrated specific heat,
is equal to $2J(q-1)/q$ for the square lattice. 
The $q$-dependence of the energy change, and therefore the vanishing
of the specific-heat amplitude at $q=1$, can thus be compensated by
introducing a normalization factor $q/(q-1)$. This is illustrated in
Fig.~\ref{fig:pcplt} which shows the specific heat of the random-cluster
model on the square lattice, including such a factor, in the limit $q \to 1$.
The quantity plotted in Fig.~\ref{fig:pcplt} is equal to
$K^2 \, {\partial ^3} f(K,q)/{\partial ^2 K}{\partial q}$ at $q=1$. 
The Monte Carlo calculation of this quantity
is slightly more involved than that of the random-cluster specific
heat \cite{QDB}  for general $q$, because of the additional derivative to
$q$, which requires sampling of the correlation of the bond density 
and the cluster density at $q=1$. In particular, our numerical results were
obtained by sampling of
\begin{displaymath}
\left( \frac{\partial ^3f(K,q)}{\partial K^2 \partial q}  \right)_{q=1}
= \left\{ \frac{(u+1)^2}{u^2}(\langle {\calN_{\rm b}}^2 {\calN_{\rm c}} / N_{\rm s} \rangle
-2\langle {\calN_{\rm b}} {\calN_{\rm c}}/N_{\rm s}\rangle \langle \calN_{\rm b} \rangle)-
\frac{u+1}{ u^2}\langle \calN_{\rm b} \calN_{\rm c}/N_{\rm s} \rangle \right.
\end{displaymath}
\begin{displaymath}
\left. - \left[
\frac{(u+1)^2}{u^2}(\langle {\calN_{\rm b}}^2\rangle -2\langle {\calN_{\rm b}} \rangle^2)-
 \frac{u+1}{u^2}\langle \calN_{\rm b} \rangle \right] \langle \calN_{\rm c}/N_{\rm s}\rangle\right\}
\end{displaymath}
and extrapolation to the thermodynamic limit. We simulated square systems
with sizes up to $L=64$, taking numbers of samples up to a few hundred
million.

The figure illustrates that the rescaled specific heat remains finite at
the critical temperature, and displays a cusp-like singularity which is,
as follows from the known temperature exponent \cite{NDL, CDL} of the Potts model,
proportional to $|T^{*}-T^{*}_{\rm c}|^{-\alpha}$, with $\alpha=-2/3$ and 
the reduced temperature $T^{*}\equiv kT/J=1/K=-1/\ln(1-p)$.

\section{Transfer-matrix calculation of the percolation connectivities} 
\label{TMcal}
The key observation behind the results of Ref.~\onlinecite{MN} is that the
leading transfer-matrix eigenvector can be normalized such that all its
components are integers. Motivated by these results, we investigated
finite $L \times \infty$ bond-percolation systems with the periodic
direction along a set of edges, for several values of $L$.
Indeed we found that it is possible to normalize the eigenvector
belonging to the largest eigenvalue such that all  components are
integers with greatest common divisor 1. While this eigenvector describes
the connectivity at the open end of the cylinder, one can connect two
of these systems by $L$ intermediate bond variables, and thus compute the
connectivities on a cylinder without an open end. It is therefore possible to
express the nearest- and the next-nearest-neighbor connectivities on these
finite systems as exact fractions.
The results of these transfer-matrix calculations
are presented in Table~\ref{TMg}.
It is apparent that the connectivities converge very quickly to their infinite-system
values $3/4$ and $11/16$ as $L$ increases.
The data were fitted by an iterated power-law method \cite{HN82}, which yielded 
$\gn=0.750\;2(8)$ and $\gnn=0.687\;49(2)$. These fit results are consistent with
the infinite-system values.

\begin{table}
\begin{center}
    \begin{tabular}{|c|c|c|c|}
    \hline 
$L$& numerator & denominator & $\gn$ \\ 
    \hline 
        $2$ & $21$ & $5^2$ & $0.84$ \\
        $3$ & $1201$ & $39^2$ & $0.789612097304$ \\
        $4$ & $1496541$ & $1393^2$ & $0.771234389567$ \\
        $5$ & $4331416849$ & $75337^2$ & $0.763156025078$ \\
        $6$ & $258134675843541$ & $18442085^2$ & $0.758972970522$ \\
        $7$ & $3885478927552013401$ & $2266262629^2$ & $0.756526392191$ \\
        $8$ & $47703428114196051853941$ & $251368505957^2$ & $0.754966814629$ \\
    \hline 

    \hline 
$L$& numerator & denominator & $\gnn$ \\ 
    \hline 
        $2$ & $16$ & $5^2$ & $0.64$ \\
        $3$ & $114$ & $13^2$ & $0.674556213018$ \\
        $4$ & $1326144$ & $1393^2$ & $0.683421208184$ \\
        $5$ & $3893316098$ & $75337^2$ & $0.685966680489$ \\
        $6$ & $233593856264336$ & $18442085^2$ & $0.686817539741$ \\
        $7$ & $3529173407855598194$ & $2266262629^2$ & $0.687151539217$ \\
        $8$ & $390852028122815173284096$ & $754105517871^2$ & $0.687302830864$ \\
    \hline 
    \end{tabular}
\end{center}
\caption{Nearest- and next-nearest-neighbor connectivities on $L \times \infty$
        square bond-percolation lattices with periodic boundary conditions
        along a set of edges \cite{note}. They are also represented as exact fractions whose
        numerators and denominators are listed.}
\label{TMg}
\end{table}

Some practical guidance is given in Ref.~\onlinecite{MN} about
how one can guess a formula from a series of integer numbers.
We did not succeed in guessing exact formulae for $\gn$ and $\gnn$
as functions of $L$. The difficulty originates from the following facts: 
(1) large prime numbers occur, such as $75337$ in the denominator 
of the fractional value of connectivities when $L=5$,
and $55051$ in the factorization of $18442085$ which occurs in the
denominator for $L=6$; and
(2) the integers in the leading eigenvector increase very rapidly as $L$ increases. 
This made clear by an inspection of the smallest elements of the leading eigenvector. 
A list of values of these smallest elements, after normalization as mentioned above,
is presented in Table~\ref{TMmin-int} for several values of $L$.

\begin{table}
\begin{center}
    \begin{tabular}{|c|c|c|c|c|c|}
    \hline 
   $L$&  2 &  4  &  6  &  8  &  10  \\
    \hline 
   $i$&  1 &  $2^2$ & $2^6$ & $2^{12}$ & $2^{20}$ \\
    \hline 
    \hline 
   $L$&  3 &  5  &  7  &  9      & \\
    \hline 
   $i$&  3 &  $15 \times 2^2$ & $63 \times 2^6$ & 
                  $255  \times 2^{12}$ & \\
    \hline 
    \end{tabular}
\end{center}
\caption{Integer value ($i$) of the smallest element in the normalized
eigenvector
which corresponds with the largest eigenvalue of the transfer-matrix for
the bond-percolation problem 
on an $L \times \infty$ square lattice with periodic boundary conditions
along a set of edges.}
\label{TMmin-int}
\end{table}

For even $L$, the entries in Table~\ref{TMmin-int} are equal to
$2^{(L-2)L/4}$ for even $L$, and for odd $L$ they are equal to
$(2^{L-1} -1) 2^{(L-1)(L-3)/4}$. Thus, defining $c_L \equiv 2^{(L-2)L/4}$,
one observes that the smallest element is $c_L$ if $L$ is even, and
$c_{L+1}-c_{L-1}$ if $L$ is odd.

Since many analytic expressions have been obtained \cite{MN} for the completely
packed $O(1)$ loop model, which relate to specific algebraic numbers series, such
as the number of symmetric alternating sign matrices and coefficients of 
the characteristic polynomial of the Pascal matrix \cite{MN},  
one wonders if it will be possible to find exact expressions for the
aforementioned connectivities as a function of $L$ in the case of the
bond-percolation problem on $L \times \infty$ square lattices with the
presently used periodic direction.

\section{Relation between percolation and $O(1)$ loop correlations} 
\label{appMapping}

\begin{figure}
\includegraphics[width=6.0cm]{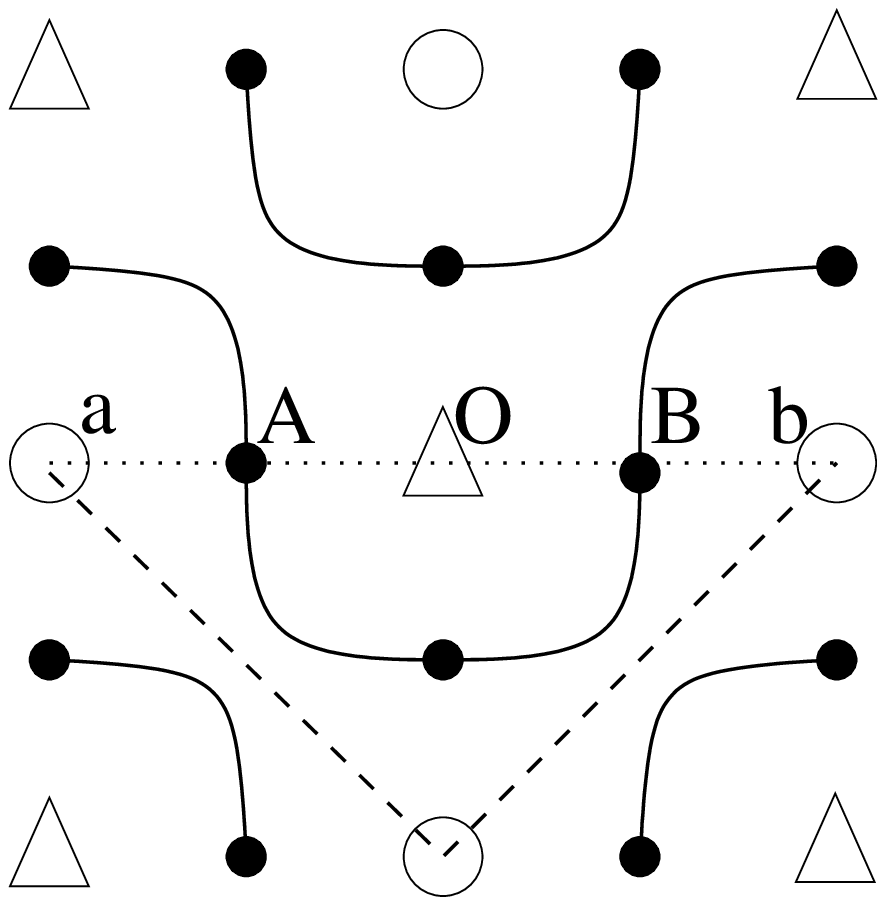}
\caption{
Correspondence between completely packed $O(1)$ loop configurations and
configurations of the bond-percolation model. The figure shows a part 
of a system wrapped on a cylinder, such that it is periodic
in the horizontal direction and extends to infinity in
both vertical directions.
Solid circles show points in the middle of the lattice edges of the $O(1)$ loop model.
The dual lattice of the lattice defined by these solid circles is divided into
two mutually dual square sublattices, 
whose lattice sites are shown by open triangles and circles.
Solid lines are for loops, and dashed lines are for bonds in percolation
clusters on one of the dual square lattices.
The dotted line indicates a row where we take the probability that
two consecutive points, such as A and B,  on a row lie on the same $O(1)$ loop,
and the probability that two next-nearest neighbor sites, such as a and b,
of the corresponding percolation configuration belong to the same cluster.}
\label{mapping}
\end{figure}

Figure \ref{mapping} illustrates the mapping of a completely
packed loop configuration to a bond configuration of the corresponding
bond-percolation problem \cite{BKW}.
Ref.~\onlinecite{MN} gives a conjecture on the probability that
$n$ consecutive points on a row lie on the same loop of the 
$O(1)$ loop model on $L \times \infty$ cylinders.
For $n=2$, it predicts that the probability approaches $11/16$ as 
$L \rightarrow \infty$. We argue that, for the completely packed  $O(1)$ loop model
on $L \times \infty$ cylinders, the probability 
that two consecutive points on a row, such as A and B in Fig.~\ref{mapping},
 lie on the same loop equals the probability that two next-nearest neighbors,
such as $a$ and $b$ in Fig.~\ref{mapping}, are in the same 
percolation cluster on the corresponding square lattice. 
The argument is based on: 
(1) When two consecutive points on a row lie on the same loop, 
the two next-nearest neighbors on the corresponding percolation lattice
belong to the same cluster. 
(2) When two consecutive points on a row lie on different loops, 
the two next-nearest neighbors on the corresponding percolation lattice
belong to different clusters.

The two conclusions above can be derived as follows.
In Fig.~\ref{mapping}, $a$ and $O$ are located on different sides of the
loop through point $A$, while $b$ and $O$ are located on different sides
of the loop through point $B$. In this configuration, $A$ and $B$ lie on
the same loop, so that $a$ and $b$ are adjacent to and on the same side of
the loop.  Therefore, $a$ and $b$ belong to the same percolation cluster
on the corresponding square lattice.
Let us now change the loop configuration such that $A$ and $B$ lie on
different loops. Then, the path $aAOBb$ crosses the loop through $A$ once, 
i.e., one of $a$ and $b$ belongs to the inside of that loop and the
other one to the outside. Therefore, $a$ and $b$ belong to different
percolation clusters.

The Mitra-Nienhuis conjecture was based on exact numerical results for
systems on a cylinder with a finite circumference, which also applies
to our transfer-matrix calculations for the percolation problem.
However, the orientation of the $O(1)$ lattice used in Ref.~\onlinecite{MN} 
with respect to the axis of the cylinder differs by $\pi/4$ from our
percolation lattice, so that our results for finite system do not
match those for the $O(1)$ model. But these differences should vanish after
extrapolation to the infinite system.

\section{Derivation of the exact nearest-neighbor connectivity for bond percolation 
on the triangular and honeycomb lattices} 
\label{gnHonDer}
We first derive the exact nearest-neighbor connectivity on the triangular
lattice, and then find the one on the honeycomb lattice using a duality relation. 
For the bond-percolation problem on the triangular lattice, 
with $-E_{\rm c}/3K_{\rm c}=1$ at $q=1$, from Eq.~(\ref{eq:tri-gnq}) one gets
\begin{equation}
 g^{\rm tri}_{\rm n,0} 
 = 1 - \left( \frac{\partial (E/3K)}{\partial q} \right)_{q=1, K=K_{\rm c}}\;.
\label{eq:tri-gn0} 
\end{equation}
The value of $K_{\rm c}$ as a function of $q$ can be obtained from 
Ref.~\onlinecite{KJ74} as 
\begin{equation}
\exp{K_{\rm c}(q)}=1+ \frac{\sqrt{q}}{2} \sec \left( \frac{1}{3} \arctan \sqrt{ \frac{4}{q} -1 } \; \right) \;, 
\label{eq:tri-Kc} 
\end{equation}
and the reduced internal energy at $K=K_{\rm c}$ is given in Ref.~\onlinecite{BTA78} as
\begin{equation}
E_{\rm c}(q) = -3 \epsilon \csc(2 \phi) \sin(2 \phi /3) \sin(4\phi/3) 
\int^{\infty}_{-\infty} \frac{\sinh[(\pi-\phi)x] \cosh(2\phi x/3)}{\sinh (\pi x) \cosh (\phi x)} {\rm d} x \;,
\label{eq:tri-E} 
\end{equation}
with $\cos \phi={\sqrt{q}}/{2} \;\;\;\; (0<\phi<\frac{\pi}{2})$, 
$\epsilon=\ln [2 \cos(2\phi/3)]$, and $q<4$.

Substituting $K_{\rm c}(q)$ and $E_{\rm c}(q)$ into Eq.~(\ref{eq:tri-gn0}), 
we derive the exact connectivity in the limit $q \rightarrow 1$ as 
\begin{eqnarray}
\gnz^{\rm tri} 
&=&1 - \frac{ 8 \ln \left(2 \cos{\frac{2\pi}{9}}\right) \sin^2{\frac{\pi}{9}} \left( 3\sqrt{3} - \tan{\frac{\pi}{9}}\right) 
\cos{\frac{\pi}{18}}}{9 \ln^2 \left[\frac{1}{2} \left( 2+\sec{\frac{\pi}{9}} \right) \right] 
\left(2+\sec{\frac{\pi}{9}}\right)} \nonumber \\
&&+ \frac{ 12\cos{\frac{\pi}{9}} -8\cos{\frac{2\pi}{9}} -4\sin{\frac{\pi}{18}} -2\ln \left( 2 \cos{\frac{2\pi}{9}} \right)
\left( 7 + 4\sin{\frac{\pi}{18}} \right)}{9 \ln \left[\frac{1}{2} \left( 2+\sec{\frac{\pi}{9}} \right) \right] 
\left(2+\sec{\frac{\pi}{9}}\right)}  \nonumber \\
&&+ \frac{ {\ln \left( 2 \cos \frac{2\pi}{9} \right)} {\cos \frac{\pi}{18} }
\left\{ -{8 \cot \frac{\pi}{9}} + {3 \csc \frac{2\pi}{9}}
+ \left[ -21+16\sqrt{13} {\cos \left(\frac{1}{3} \arctan{\frac{53\sqrt{3}}{19}}\right)}\right]
{\sin \frac{2\pi}{9}} \right\} }
{9 \ln \left[\frac{1}{2} \left( 2+\sec{\frac{\pi}{9}} \right) \right] } \nonumber \\
&=&\frac{3(2+7 p_c^{\rm tri})}{4(5-p_c^{\rm tri})}
=\frac{3(2+14\sin[\pi/18])}{4(5-2\sin[\pi/18])}
=0.714\;274\;133\;\cdots\;,
\label{tri-gn-exa}
\end{eqnarray}
where $p_c^{\rm tri}=2\sin(\pi/18)$ is the bond-percolation threshold on the triangular 
lattice \cite{SE64}.
The derivation involved the calculation of several complicated integrals, 
which led to an intermediate result (the first three lines of Eq.~(\ref{tri-gn-exa})).
We found the simplified expression in the last line of Eq.~(\ref{tri-gn-exa}) with 
the help of an answer engine \cite{WA} using numerical values of
the intermediate result. We verified that the two results are exactly equal. 
In the verification, we made use of the identities 
$4 \cos(2 \pi/9) = 2 + \sec(\pi/9)$ 
and 
$\sqrt{13} \cos(\arctan[{53\sqrt{3}}/{19}]/3) = (4+7\sin[\pi/18])/(3-8\sin[\pi/18])$. 

From the above $\gnz^{\rm tri}$, one obtains the value of $\gnz^{\rm hon}$ as follows. 
Let  $p^{\rm tri}_{\rm c}$ be the critical bond-occupation probability on the triangular
lattice, and $p^{\rm tri}_{\rm o}$ the probability that two nearest-neighbor sites are
connected via some path of bonds not covering the bond between the two sites.
Then, $(1-p^{\rm tri}_{\rm c}) p^{\rm tri}_{\rm o}$ is the probability that there is no
bond between nearest-neighbor sites, while the sites are still connected. Thus
\begin{eqnarray}
\gn^{\rm tri}=p^{\rm tri}_{\rm c} + (1-p^{\rm tri}_{\rm c}) p^{\rm tri}_{\rm o} \;.
\label{tri-gn-i}
\end{eqnarray}
Similarly, one can write for the honeycomb lattice 
\begin{eqnarray}
\gn^{\rm hon}=p^{\rm hon}_{\rm c} + (1-p^{\rm hon}_{\rm c}) p^{\rm hon}_{\rm o} \;.
\label{hon-gn-i}
\end{eqnarray}
The duality property tells that 
\begin{eqnarray}
p^{\rm tri}_{\rm c}+ p^{\rm hon}_{\rm c}=1 \;,\;\;\;\;
p^{\rm tri}_{\rm o}+ p^{\rm hon}_{\rm o}=1 \;.
\label{duality-relations}
\end{eqnarray}
The substitution of Eqs.~(\ref{duality-relations}) and (\ref{tri-gn-i})
into Eq.~(\ref{hon-gn-i}) yields 
\begin{equation}
\gn^{\rm hon}=
1-p_c^{\rm tri}+p_c^{\rm tri} 
\left(1-\frac{\gn^{\rm tri}-p_c^{\rm tri}}{1-p_c^{\rm tri}}\right) 
\label{gn-hon}
\end{equation}
Using the above equation and the $\gnz^{\rm tri}$ value as given 
in Eq.~(\ref{tri-gn-exa}), one obtains 
\begin{eqnarray}
\gnz^{\rm hon}
&=& \frac{3 (7-9p_c^{\rm tri})}{4(5-4p_c^{\rm tri})}
= \frac{3 (-2+9p_c^{\rm hon})}{4(1+4p_c^{\rm hon})}
= \frac{3(-7+18\sin[\pi/18])}{4(-5+8\sin[\pi/18])} \nonumber \\
&=& 0.804\;735\;202\;\cdots.
\label{hon-gn-exa}
\end{eqnarray}


\begin{thebibliography}{100}
\bibitem{note}
The $g_n$ values presented in this table was calculated 
in the vertical direction of the cylinder (the direction which extends to infinity). 
We also conducted an exact enumeration for $g_n$ in the horizontal direction 
with $2 \le L \le 8$. These results tells that the average of $g_n$ in the two directions 
is exactly $3/4$ for $2 \le L \le 8$. 
By using a self-dual property of the square lattice, it can be exactly proved that
the average of $g_n$ in the two directions is $3/4$ for $L \ge 2$.
\end{thebibliography}
\end{document}